\documentstyle[twoside,fleqn,knuckle]{article}
\newcommand{\ncm}{\newcommand}
\ncm{\nrml}{\normalsize}
\ncm{\be}{\begin{equation}}
\ncm{\ee}{\end{equation}}
\ncm{\bea}{\begin{eqnarray}}
\ncm{\eea}{\end{eqnarray}}
\ncm{\bedm}{\begin{displaymath}}
\ncm{\eedm}{\end{displaymath}}
\ncm{\dty}{\displaystyle}
\ncm{\tsy}{\textstyle}
\ncm{\ssy}{\scriptstyle}
\ncm{\sssy}{\scriptscriptstyle}
\ncm{\dg}{\dagger}
\ncm{\hlf}{{\tsy \frac{1}{2}}}
\ncm{\qrt}{{\tsy \frac{1}{4}}}
\ncm{\thrqrt}{{\tsy  \frac{3}{4}}}
\ncm{\lsim}{\lower.7ex\hbox{$\,\stackrel{\tsy <}{\sim}\,$}}
\ncm{\gsim}{\lower.7ex\hbox{$\,\stackrel{\tsy >}{\sim}\,$}}
\ncm{\htm}{\hat{\mu}}
\ncm{\htn}{\hat{\nu}}
\ncm{\htr}{\hat{\rho}}
\ncm{\dau}{\partial_{\mu}}
\ncm{\dauu}{\partial_{\nu}}
\ncm{\Dl}{\Delta}
\ncm{\dl}{\delta}
\ncm{\bt}{\beta}
\ncm{\gm}{\gamma}
\ncm{\th}{\theta}
\ncm{\ta}{\tau}
\ncm{\sg}{\sigma}
\ncm{\ep}{\varepsilon}
\ncm{\nn}{\nonumber}
\ncm{\ra}{\rightarrow}
\ncm{\xtl}{\tilde{x}}
\ncm{\btc}{\beta_c}
\ncm{\bts}{\beta_{\mbox{\small s}}}
\ncm{\btg}{\beta_{\mbox{\small g}}}
\ncm{\CS}{C_{\mbox{\small s}}}
\ncm{\CD}{C_{\mbox{\small d}}}
\ncm{\cs}{C_{\mbox{s}}}
\ncm{\cd}{C_{\mbox{d}}}
\input{psfig}
\title{The Behavior of Vortex Loops in the 3-d XY Model
\\[0.4\baselineskip]
\vspace*{-25mm}
{\normalsize\noindent November 1993 \hfill LTH 323 \\
\mbox{} \hfill hep-lat/9311042 \\}
\vspace*{17mm}
}
\author{{\large Arjan Hulsebos}\thanks{Present address: FB 8, University of
Wuppertal, Gau\ss stra\ss e 20, 42119 Wuppertal, BRD.}
\thanks{email: {\tt arjanh@wpts0.physik.uni-wuppertal.de} }
\\[0.5\baselineskip]
DAMTP, Chadwick Tower, University of Liverpool, P.~O.~Box 147,
Liverpool, L69 3BX, UK. }
\begin{document}
\begin{abstract}
The behavior of vortex loops is studied in the 3-d XY model. It is found that
the phase transition of the 3-d XY model is caused by percolating vortex loops.
\vspace*{-1\baselineskip}
\end{abstract}
\maketitle

The 3-d XY model is known to allow for vortices and antivortices. If we
impose periodic boundary conditions, these vortices and antivortices have
to form closed loops. Since the 3-d XY model undergoes a phase transition
at finite coupling $\btc \approx 0.4542$ \cite{GotHas93}, we may suspect
that those loops behave differently on either side of the phase transition.

Here, we will study the behavior of these loops by defining suitble
correlation functions and loop distributions. Our conclusion is that the
phase transition coincides with a percolation transition for the vortex loops.
\vspace*{\baselineskip}

The action for XY models is given by
\be
S = \bt \;\sum_{x,\mu}\; \cos(\th_{x+\htm}-\th_x),\;\;\mu= 1,2,\ldots,d.
\label{XYact}
\ee
The quantity
\bedm
k_{\mu\nu}(x) = \frac{1}{2\pi} \Bigl\{ [\th_{x+\htm} - \th_x] +
 [\th_{x+\htm+\htn} - \th_{x+\htm}]
\eedm
\vspace*{-1.1\baselineskip}
\be
 \;\;\; + \; [\th_{x+\htn} - \th_{x+\htm+\htn}] + [\th_x - \th_{x+\htn}]
\Bigr\},\;\; \mu < \nu, \label{vortex}
\ee
where $[\ldots ]$ denotes the restriction to the interval $<-\pi,\pi]$,
yields the vortex number on the corresponding plaquette. Turning now to 3-d,
we perform a dual transformation by defining
\be
j_{\mu}(\xtl-\htm) = \hlf \ep_{\mu\nu\rho}k_{\nu\rho}(x), \label{vlink}
\ee
where $\xtl$ on the dual lattice is identified with $x+\hlf\htm+\hlf\htn+
\hlf\htr$ on the original lattice.

\begin{figure}
\vspace*{-2.2cm}
\hspace*{-1.0cm}
\centerline{%
    \hbox{%
    \psfig{figure=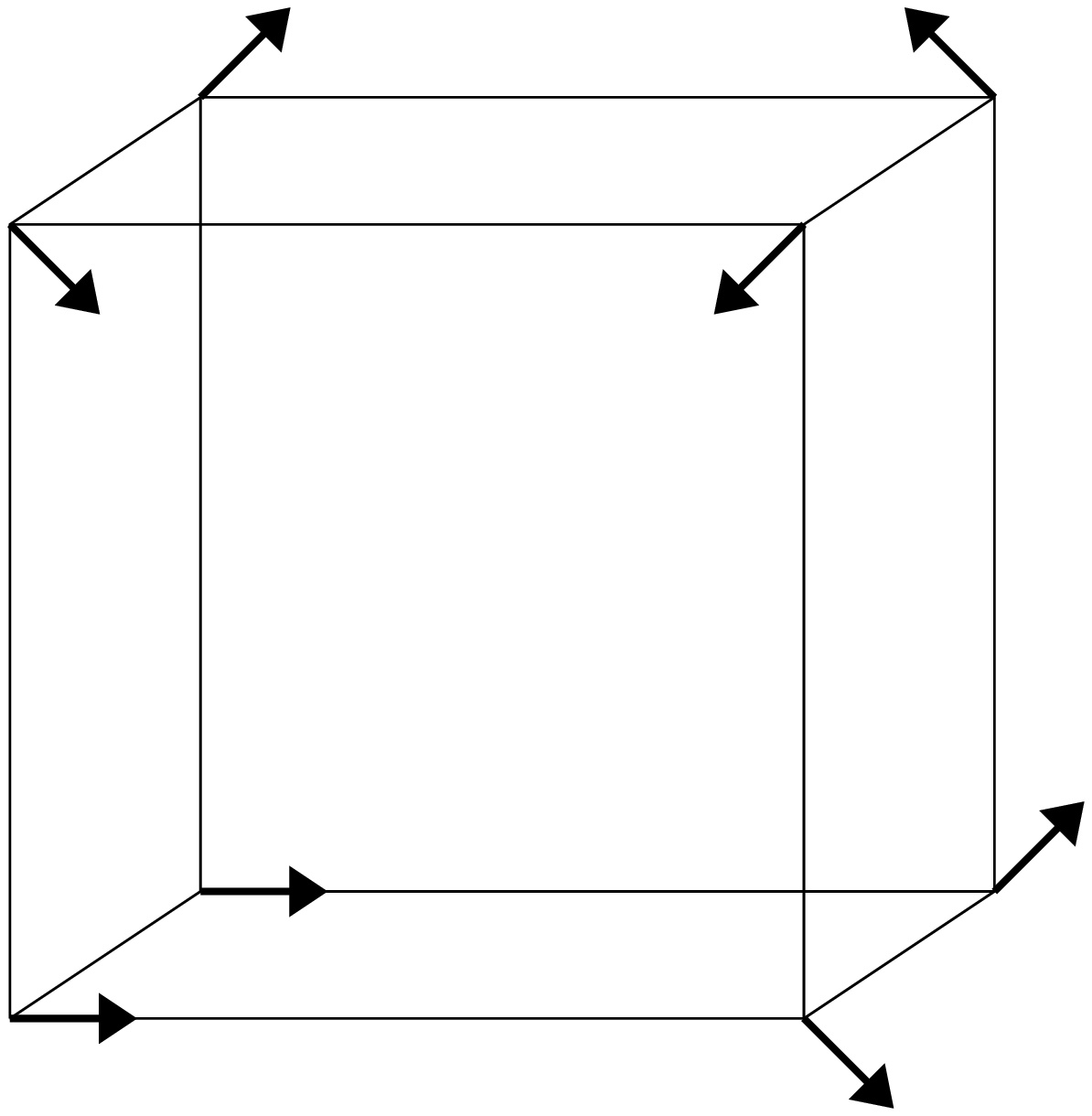,height=3in,width=3in}%
   }%
}
\vspace{-1cm}
\caption{Spin configuration as given in (\protect\ref{spins}). The spins are in
the 1-3 plane. }
\label{cube}
\vspace*{-0.5cm}
\end{figure}

The links $j_{\mu}$ have to satisfy the following relation, dropping the tilde
from now on,
\bea
\dau' j_{\mu}(x) & = & 0 , \label{j-conserv} \\
\dau'f(x) & \equiv & f(x) - f(x - \htm)  \nn
\eea
This means that each point $x$ on the dual lattice has to be visited by an
even number of non-zero links $j$. Hence the loops have to be closed on a
finite lattice.  To such a conservation law one usually coins the phrase
``what comes in, must go out again'', but this need not be true on a lattice.
Consider the following spin configuration on an elementary cube:
\bea
& & \th_x = \th_{x+\hat{2}} = 0,
\;\;\th_{x+\hat{1}} = \th_{x+\hat{3}} = -\qrt\pi, \nn \\
& & \th_{x+\hat{1}+\hat{3}} =  -\thrqrt\pi, \;\;
    \th_{x+\hat{1}+\hat{2}+\hat{3}} = \thrqrt\pi , \nn \\
& & \th_{x+\hat{2}+\hat{3}} = \th_{x+\hat{1}+\hat{2}} = \qrt\pi.
\label{spins}
\eea
See also figure \ref{cube}. This configuration will lead to a vortex ($k=1$) on
the right face, and an antivortex ($k=-1$) on the top face. The $\ep$-tensor
conspires in such a way as to have the same sign for both non-zero $k$'s,
leading to
\bea
& & j_1(x) = 1,\; j_2(x) = 0,\; j_3(x) = -1, \nn \\
& & j_1(x-\hat{1}) = j_2(x-\hat{2}) = j_3(x-\hat{3}) = 0.
\label{currs}
\eea
So we see that ``what comes in, doesn't go out, and what goes out, doesn't come
in''! However, relation (\ref{j-conserv}) remains satisfied for this spin
configuration. This also implies that it does not make much sense to
distinguish between vortex loops and antivortex loops, since any loop may
consist of links $j=1$ as well as of links $j=-1$. We shall therefore make no
distinction and call all loops `vortex loops'. In order to study these loops,
we define two correlation functions:
\bea
\CS(r) & = &
\sum_{\stackrel{\ssy x,y,i}{\ssy |x-y|=r}} \dl_{x\in l_i} \dl_{y\in l_i}
\Bigl/ \Bigr. \sum_{\stackrel{\ssy x,y}{\ssy |x-y|=r}}
\hbox{\LARGE\lower.5ex\hbox{1\nrml ,}} \label{Cs} \\
\CD(r) & = &
\sum_{\stackrel{\ssy x,y,i,j}{\ssy |x-y|=r}} \dl_{x\in l_i} \dl_{y\in l_j}
\Bigl/  \Bigr. \sum_{\stackrel{\ssy x,y}{\ssy |x-y|=r}}
\mbox{\LARGE\lower.5ex\hbox{1}} \nn \\
& & \;\;\;\;\;\;\; - \Bigl(\frac{1}{V} \sum_{x,i} \dl_{x\in l_i}\Bigr)^2 .
\label{Cd}
\eea
(\ref{Cs}) is the probability that two points $x$ and $y$, a distance $r$ away,
are on the same loop $l_i$, and was already studied in \cite{PPY91}.

The simulations were done using a 10-hit Metropolis algorithm with site
dependent stepsize, combined with $\gm=2$ multigrid, as described in
\cite{HSV91}.  We made runs consisting of 10 batches of 100 measurements each,
after discarding 500 measurements for thermalization.

We made runs for $\bt = 0.00$ up to $\bt = 1.00$ in steps of $\Dl\bt = 0.05$,
as well as runs with $\Dl\bt = 0.01$ for $\bt$ between 0.40 and 0.50.

\begin{figure}
\vspace*{-1.2cm}
\hspace*{-1.0cm}
\centerline{%
    \hbox{%
    \psfig{figure=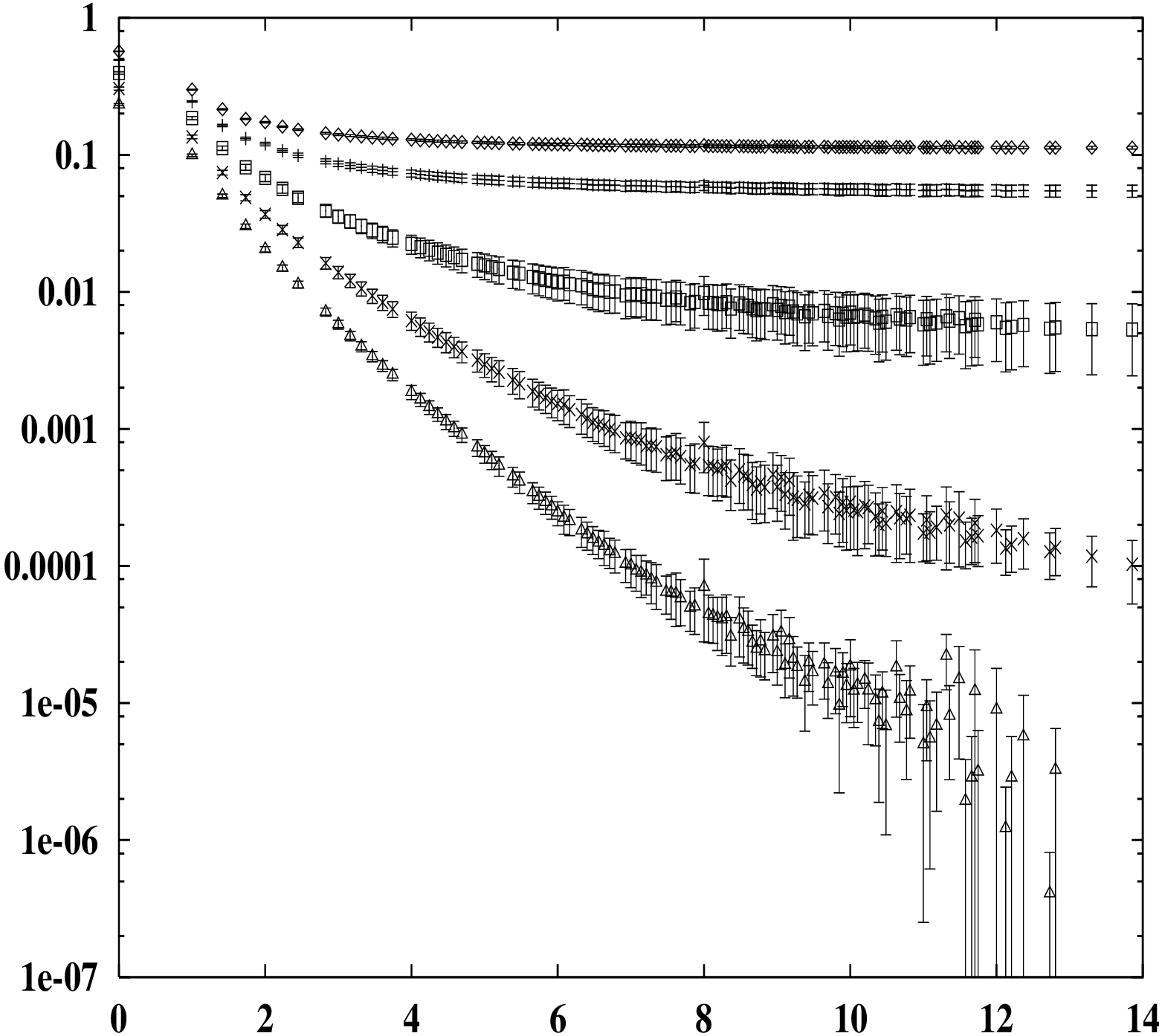,height=150pt,width=150pt}%
   }%
}
\vspace*{-0.7cm}
\caption{$\cs(r)$ for $\bt = 0.40$ (upper), $\bt = 0.44$, $0.46$, $0.48$ and
$\bt = 0.50$ (lower) on $16^3$.}
\label{Cspic}
\vspace*{-0.5cm}
\end{figure}

The results for $\CS(r)$ for $16^3$ are given in figure \ref{Cspic}. The errors
on $\CS$ and $\CD$ were obtained by averaging over the 10 batches.

As can be seen, there is a dramatic change in $\CS$ around the phase
transition.

\begin{figure}[h]
\vspace*{-1.2cm}
\hspace*{-2.0cm}
\centerline{%
    \hbox{%
    \psfig{figure=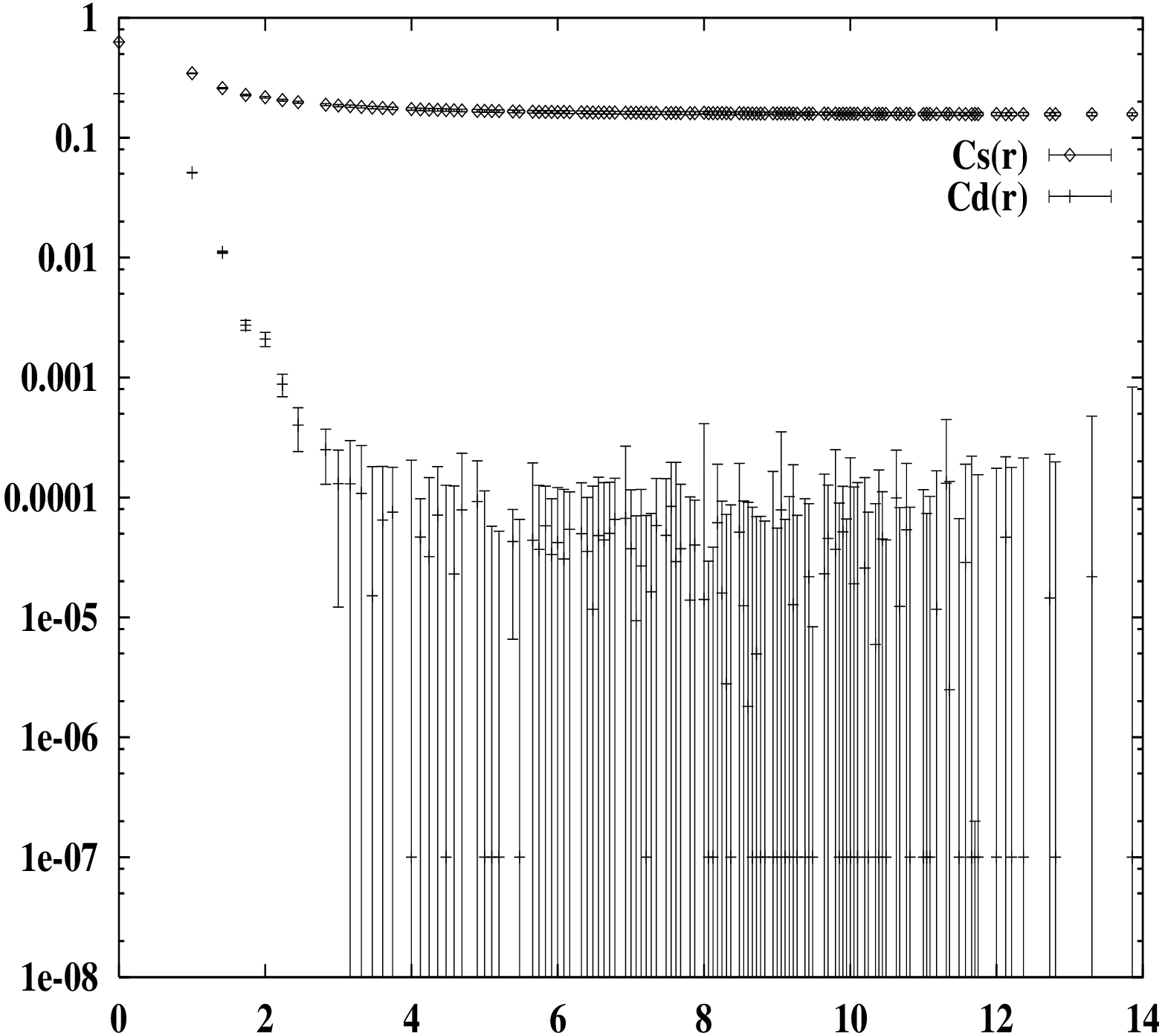,height=150pt,width=150pt}%
   }%
}
\vspace*{-4mm}
\centerline{%
    \hbox{%
    \psfig{figure=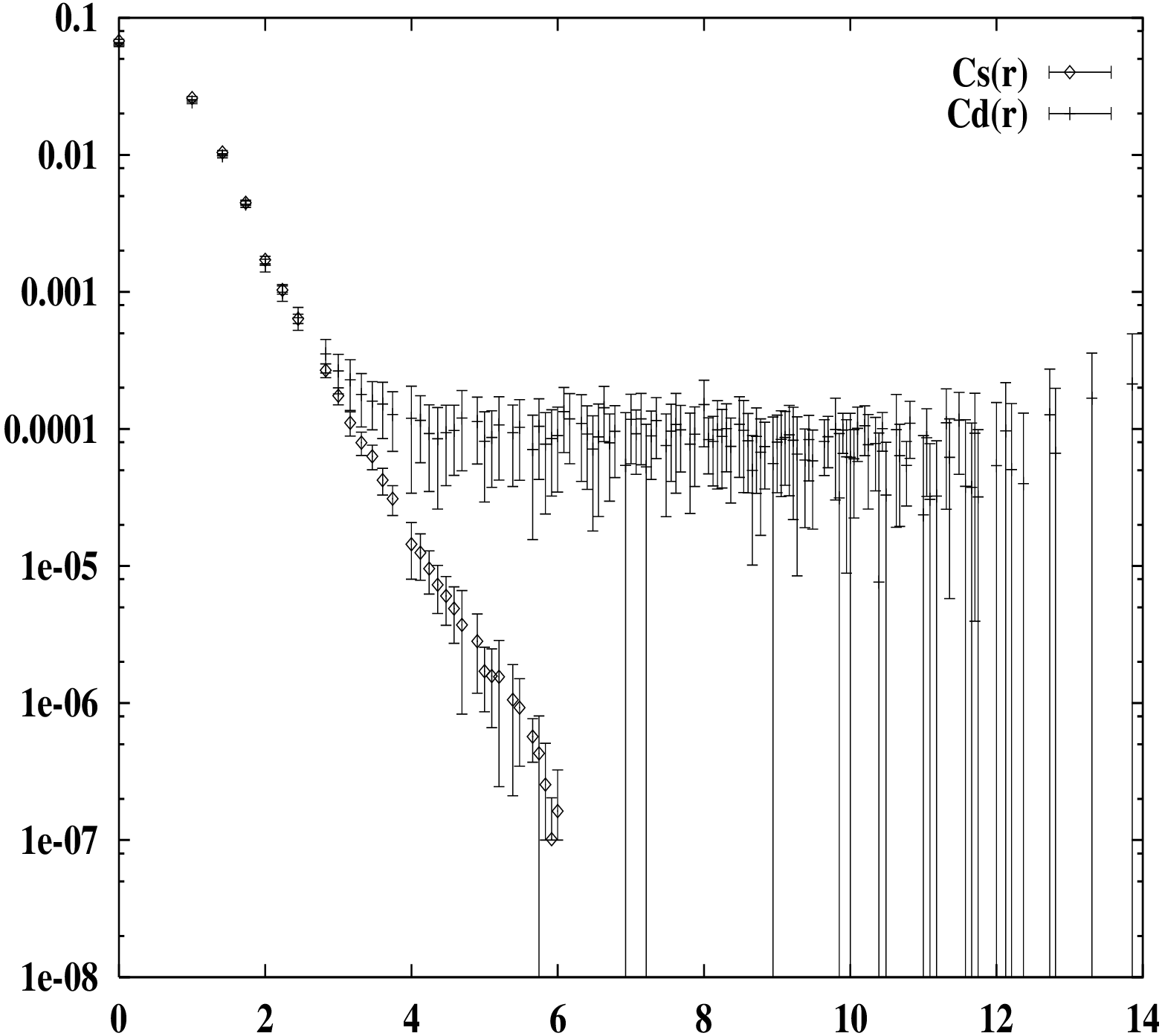,height=150pt,width=150pt}%
   }%
}
%\vspace*{-1.7cm}
\caption{$\cs$ and $\cd$ on $16^3$ for $\bt = 0.35$ (top) and for $\bt = 0.60$
(bottom) }
\label{cs-vs-cd}
\vspace*{-0.5cm}
\end{figure}

To determine whether the different vortex loops interact with one another, we
have plotted $\CS$ and $\CD$ in figures \ref{cs-vs-cd}a and \ref{cs-vs-cd}b.
We see that $\CD$ falls off exponentially for both $\bt$-values, and for
$\bt = 0.60$ differs from $\CS$ only by a constant.

In table \ref{tabone}, we present the results from fitting $\CS$ to
\be
\;\;\;\;\CS = d\;\frac{\exp(-a\,r)}{r^b} \;+\; c
\label{fitform}
\ee
for $\bt$-values around the critical coupling $\btc$.

\begin{table}[h]
\centering
\begin{tabular}{c|cccc} \hline
$\bt$ & $a$ & $b$ & $c$ & $d$ \\ \hline
& & & & \\[-1.4\baselineskip]
0.40 & $0.09^{+0.04}_{-0.04}$             & $1.56^{+0.16}_{-0.14}$
     & $0.112^{+0.002}_{-0.002}$           & $0.204^{+0.015}_{-0.015}$ \\[1mm]
0.41 & $0.10^{+0.05}_{-0.05}$             & $1.5^{+0.2}_{-0.2}$
     & $0.098^{+0.004}_{-0.003}$           & $0.208^{+0.015}_{-0.016}$ \\[1mm]
0.42 & $0.12^{+0.10}_{-0.08}$             & $1.5^{+0.4}_{-0.3}$
     & $0.089^{+0.004}_{-0.005}$           & $0.21^{+0.03}_{-0.03}$ \\[1mm]
0.43 & $0.12^{+0.07}_{-0.06}$             & $1.5^{+0.2}_{-0.3}$
     & $0.070^{+0.005}_{-0.005}$           & $0.21^{+0.03}_{-0.02}$ \\[1mm]
0.44 & $0.10^{+0.07}_{-0.07}$             & $1.4^{+0.3}_{-0.2}$
     & $0.053^{+0.005}_{-0.005}$           & $0.21^{+0.03}_{-0.03}$ \\[1mm]
0.45 & $0.08^{+0.18}_{-0.10}$             & $1.5^{+0.5}_{-0.3}$
     & $0.032^{+0.007}_{-0.008}$           & $0.20^{+0.03}_{-0.03}$ \\[1mm]
0.46 & $0.21^{+0.09}_{-0.07}$             & $1.3^{+0.3}_{-0.3}$
     & $0.005^{+0.003}_{-0.003}$           & $0.22^{+0.04}_{-0.04}$ \\[1mm]
0.47 & $0.34^{+0.06}_{-0.05}$             & $1.22^{+0.19}_{-0.15}$
     & $0.0008^{+0.0007}_{-0.0008}$        & $0.22^{+0.03}_{-0.03}$ \\[1mm]
0.48 & $0.45^{+0.06}_{-0.04}$             & $1.26^{+0.21}_{-0.16}$
     & $0.00012^{+0.00012}_{-0.00013}$     & $0.21^{+0.04}_{-0.04}$ \\[1mm]
0.49 & $0.59^{+0.06}_{-0.03}$             & $1.25^{+0.14}_{-0.12}$
     & $0.00001^{+0.00003}_{-0.00002}$     & $0.21^{+0.03}_{-0.02}$ \\[1mm]
0.50 & $0.70^{+0.04}_{-0.04}$             & $1.35^{+0.15}_{-0.12}$
     & ---                                & $0.21^{+0.05}_{-0.04}$ \\[1mm]
\hline
\end{tabular}
\caption{The results from fitting $\cs$ to (\protect\ref{fitform}) for $16^3$.}
\label{tabone}
\end{table}

\begin{figure}[h]
\vspace*{-1.2cm}
\hspace*{-1.0cm}
\centerline{%
    \hbox{%
    \psfig{figure=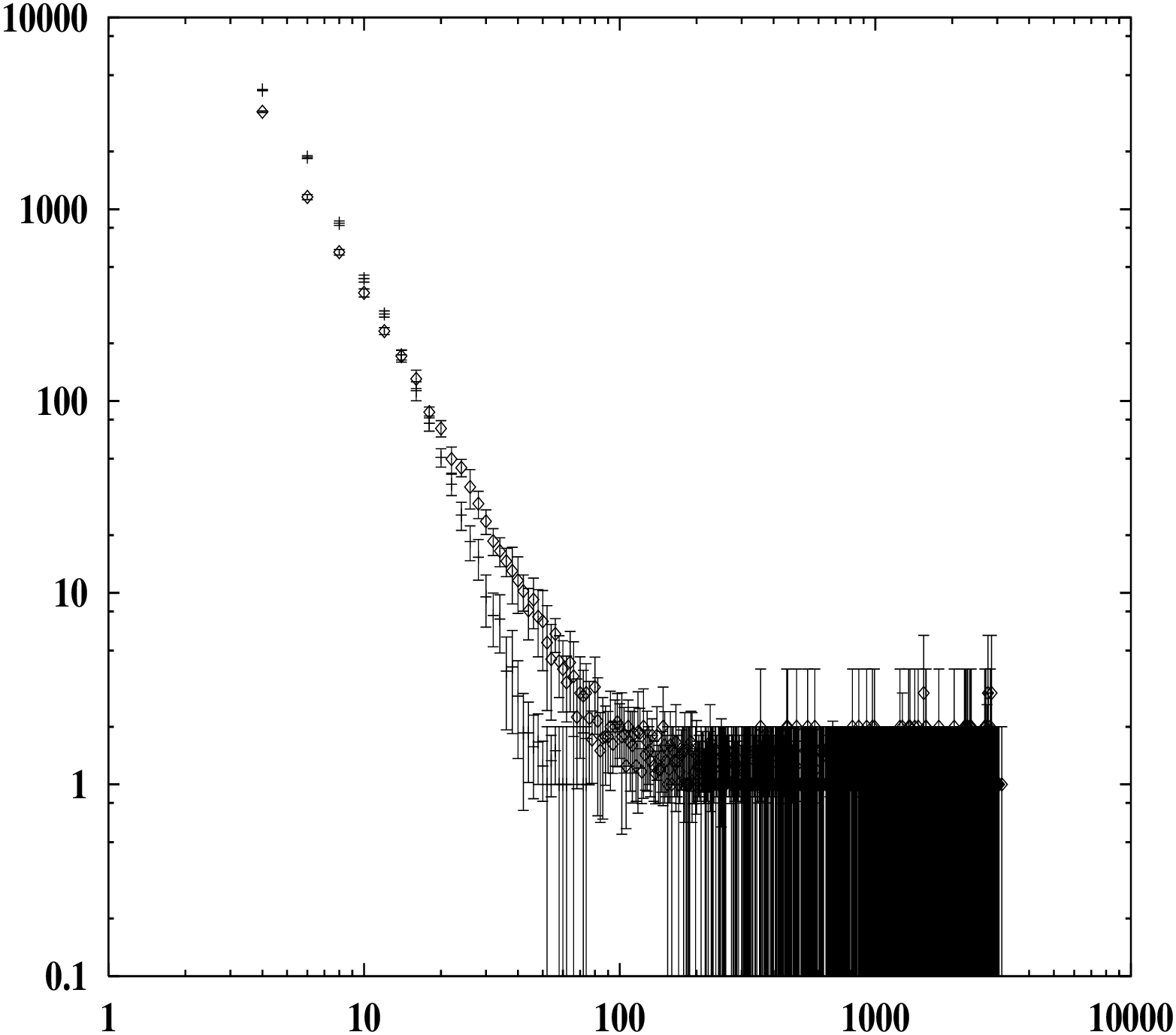,height=150pt,width=150pt}%
   }%
}
\vspace*{-0.7cm}
\caption{The number of loops of length $l$ against $l$, from $16^3$, on a
double log plot, for $\bt=0.25$ and $\bt=0.55$.}
\label{looplength}
\vspace*{-0.5cm}
\end{figure}

\begin{figure}[h]
\vspace*{-1.2cm}
\hspace*{-1.0cm}
\centerline{%
    \hbox{%
    \psfig{figure=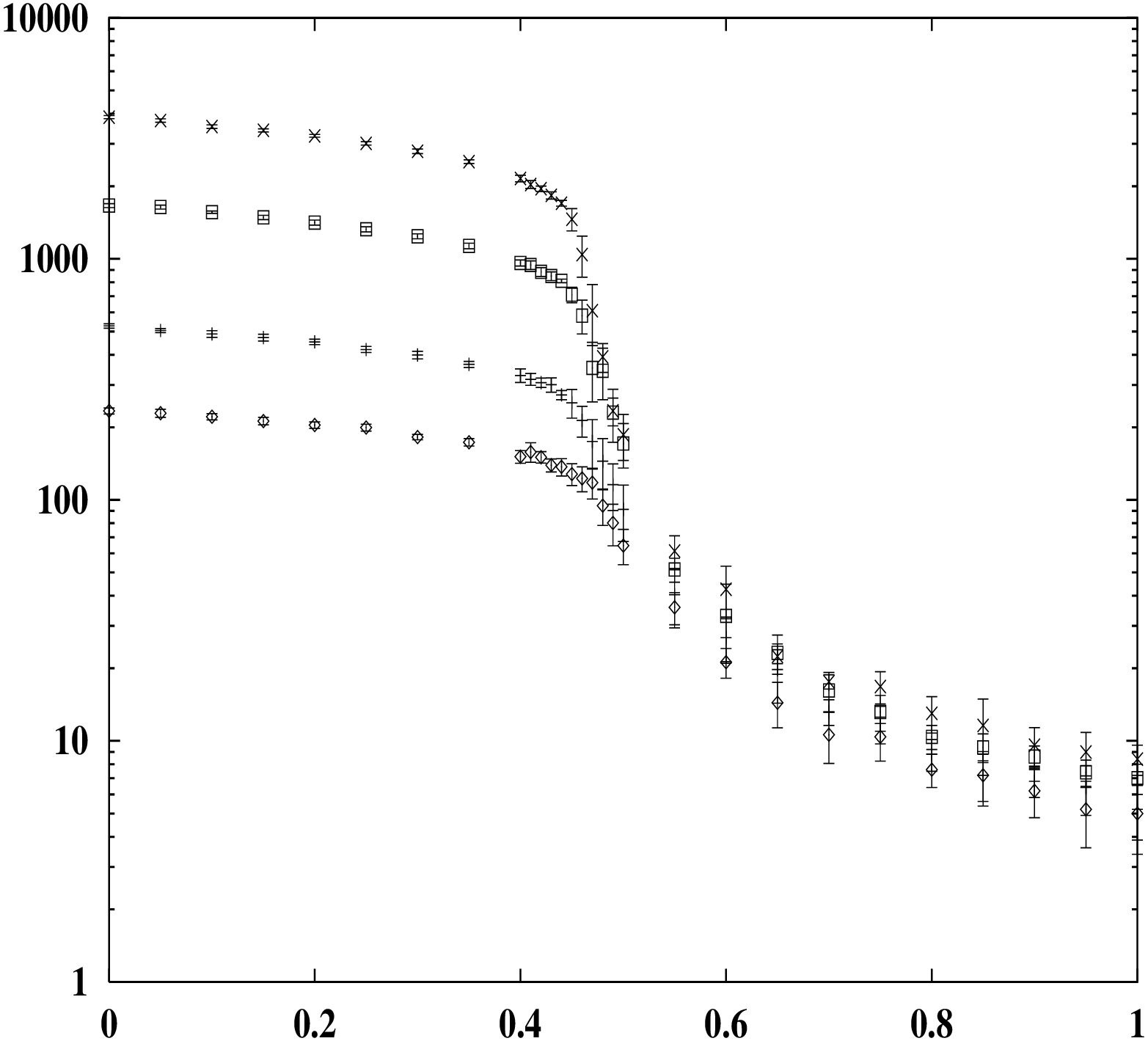,height=150pt,width=150pt}%
   }%
}
\vspace*{-0.7cm}
\caption{The average length of the longest loop against $\bt$ on a single log
plot. The lattice sizes range from $6^3$ (lower), to $16^3$ (top).}
\label{longestloop}
\vspace*{-0.5cm}
\end{figure}

Turning our attention to the length of the vortex loops, in figure
\ref{looplength} we have displayed the histogram of the length of the vortex
loops on a $16^3$ lattice for $\bt = 0.25$ and $\bt = 0.55$. Here, we also see
different behavior on either side of the phase transition.
For $\bt = 0.25$, and indeed for all $\bt\lsim\btc$, we see that there are very
long loops  present in the system. For $\bt = 0.55$, and more general for
$\bt\gsim\btc$, we see that these very long loops have vanished, and the
longest loops measure only a few times the lattice extension. This is shown in
figure \ref{longestloop}, in which the average length of the longest loop is
plotted against $\bt$, for all four lattice sizes used.

\vspace*{\baselineskip}

These results can be explained in the following way. For $\bt<\btc$, the vortex
loops may be very long, and they percolate through the lattice. In that region,
$\CS$ will tend to a constant for large separations $r$. This constant is
related to the density of percolating loops. As $\bt$ approaches $\btc$, this
density decreases, and becomes very small at the phase transition. As the
percolating loops disorder the system, the only correlation length for these
loops is the lattice size, which causes the small exponential factor $a$ in
table \ref{tabone}. The constants $a, b$ and $d$ appear to be independent
of $\bt$.

For $\bt>\btc$, the relevant degrees of freedom are the spin waves, while only
short vortex loops are allowed. As these loops have a predominatly planar
structure, we see that $a$ increases with $\bt$, while $b$ tends to zero, and
it vanishes for $\bt\gg\btc$.

There is no sign for interactions between vortex loops as $\CD$ falls off
exponentially, and differs by a constant from $\CS$ for $\bt<\btc$.
This is hardly surprizing, as loops may consist of links $j=1$ as well as
links $j=-1$.

Our results indicate that the phase transition in the 3-d XY model is caused
by percolating, non-interacting vortex loops.

\section*{ACKNOWLEDGEMENTS}

\mbox{} \vspace*{-0.5mm}
We would like to thank Chris Michael, Jan Smit and Peer Ueberholz for useful
discussions.  This work was supported by EC contract SC1 *CT91-0642.
\hspace*{-0.5\baselineskip}

\end{document}